\documentclass[runningheads]{llncs}

\usepackage{placeins} 

\usepackage{adjustbox} 

\usepackage{stfloats}

\usepackage[colorinlistoftodos,prependcaption]{todonotes} 
\presetkeys%
    {todonotes}%
    {inline,backgroundcolor=orange}{}
    
\usepackage{fontawesome}
\usepackage[nopostdot,acronym,shortcuts,nonumberlist]{glossaries}
\newacronym{(S+N)/NR}{(S+N)/NR}{signal-plus-noise-to-noise ratio}
\newacronym{ADC}{ADC}{analog-to-digital converter}
\newacronym{BCI}{BCI}{bulk current injection}
\newacronym{CAN}{CAN}{control area network}
\newacronym{CRT}{CRT}{cathode ray tube}
\newacronym{CSMACD}{CSMA/CD}{carrier sense multiple access with collision detection}
\newacronym{CoE}{CoE}{CAN over Ethernet}
\newacronym{DAC}{DAC}{digital-to-analog converter}
\newacronym{DoS}{DoS}{Denial of Service}
\newacronym{EMR}{EMR}{electromagnetic radiation}
\newacronym{FCS}{FCS}{frame check sum}
\newacronym{FEC}{FEC}{forward error correction}
\newacronym{FPGA}{FPGA}{field-programmable gate array}
\newacronym{ICMP}{ICMP}{Internet control message protocol}
\newacronym{IEC}{IEC}{International Electrotechnical Commission}
\newacronym{MAC}{MAC}{Medium Access Control}
\newacronym{NIC}{NIC}{network interface card}
\newacronym{NSA}{NSA}{National Security Agency}
\newacronym{PA}{PA}{power amplifier}
\newacronym{PC}{PC}{personal computer}
\newacronym{RF}{RF}{radio frequency}
\newacronym{SDR}{SDR}{software-defined radio}
\newacronym{SFD}{SFD}{start frame delimiter}
\newacronym{SNR}{SNR}{signal-to-noise ratio}
\newacronym{USRP}{USRP}{Universal Software Radio Peripheral}
\newacronym{SSP}{SSP}{Secure Simple Pairing}
\newacronym{USB}{USB}{Universal Serial Bus}
\newacronym{L2CAP}{L2CAP}{Logical Link Control and Adaptation Protocol}
\newacronym{AOP}{AOP}{Always-On Processor}
\newacronym{ACL}{ACL}{Asynchronous Connec\-tion-Less}
\newacronym{BLE}{BLE}{Bluetooth Low Energy}
\newacronym{BR}{BR}{Basic Rate}
\newacronym{CID}{CID}{Channel ID}
\newacronym{PDU}{PDU}{Protocol Data Unit}
\newacronym{B-Frame}{B-Frame}{Basic Information Frame}
\newacronym{C-Frame}{C-Frame}{Control Frame}
\newacronym{PSM}{PSM}{Protocol/Service Multiplexer}
\newacronym{HCI}{HCI}{Host Controller Interface}
\newacronym{EDR}{EDR}{Enhanced Data Rate}
\newacronym{SCO}{SCO}{Synchronous Connection-Oriented}
\newacronym{HSP}{HSP}{Headset Profile}
\newacronym{ATT}{ATT}{Attribute Protocol}
\newacronym{TCP}{TCP}{Transmission Control Protocol}
\newacronym{SoC}{SoC}{System-on-a-Chip}
\newacronym{IoT}{IoT}{Internet of Things}
\newacronym{GPS}{GPS}{Global Positioning System}
\newacronym{CERT}{CERT}{Computer Emergency Response Team}
\newacronym{API}{API}{Application Programming Interface}
\newacronym{MITM}{MITM}{Machine-in-the-Middle}
\newacronym{TLS}{TLS}{Transport Layer Security}
\newacronym{OTA}{OTA}{Over-the-Air}
\newacronym{XSS}{XSS}{Cross-Site Scripting}
\newacronym{SDK}{SDK}{Software Development Kit}
\newacronym{MQTT}{MQTT}{Message Queuing Telemetry Transport}
\newacronym{CVE}{CVE}{Common Vulnerabilities and Exposures}
\newacronym{ECDH}{ECDH}{Elliptic-curve Diffie–Hellman}
\newacronym{AEAD}{AEAD}{Authenticated Encryption with Associated Data}
\newacronym{AES}{AES}{Advanced Encryption Standard}
\newacronym{TOR}{TOR}{The Onion Router}
\newacronym{GDPR}{GDPR}{EU General Data Protection Regulation}
\newacronym{HSTS}{HSTS}{HTTP Strict Transport Security}
\newacronym{AWS}{AWS}{Amazon Web Services}
\newacronym{LTK}{LTK}{Long Term Key}
\newacronym{LK}{LK}{Link Key}
\newacronym{RCE}{RCE}{Remote Code Execution}
\newacronym{SIV}{SIV}{Synthetic Initialization Vector}
\newacronym{PoC}{PoC}{Proof of Concept}
\newacronym{XPC}{XPC}{Cross-Process Communication}
\newacronym{MFi}{MFi}{Made for iPhone/iPad/iPod}
\newacronym{ACI}{ACI}{Apple Controller Interface}
\newacronym{ECB}{ECB}{Electronic Codebook}
\newacronym{UART}{UART}{Universal Asynchronous Receiver-Transmitter}
\newacronym{RSSI}{RSSI}{Received Signal Strength Indicator}
\newacronym{BCS}{BCS}{Bluetooth Core Scheduler}
\newacronym{ARI}{ARI}{Apple Remote Invocation}
\newacronym{QMI}{QMI}{Qualcomm MSM Interface}
\newacronym{TLV}{TLV}{Type Length Value}
\newacronym{RIL}{RIL}{Radio Interface Layer}
\newacronym{ISTP}{ISTP}{Intel System Trace Protocol}
\newacronym{AWDL}{AWDL}{Apple Wireless Direct Link}
\newacronym{ASLR}{ASLR}{Address Space Layout Randomization}
\newacronym{MDS}{MDS}{Multi-dimensional Scaling}

\tikzstyle{line} = [draw, -latex']
\usetikzlibrary{shapes,arrows,calc,positioning,matrix}
\usetikzlibrary{positioning}
\tikzset{
    >=triangle 45
}

\usepackage[binary-units]{siunitx}
\usepackage{tabularx}

\definecolor{darkred}{rgb}{0.831, 0, 0.063}
\definecolor{darkgreen}{RGB}{11,196,1}
\definecolor{darkpurple}{RGB}{158,0,160}
\definecolor{sorange}{rgb}{0.95, 0.57, 0}
\definecolor{darkergreen}{RGB}{6,98,1}
\definecolor{darkblue}{RGB}{0,73,218}
\definecolor{frida}{RGB}{239,100,86}

\colorlet{orange}{sorange}
\colorlet{sblue}{darkblue}
\colorlet{blue}{sblue}

\usepackage{listings}
\lstdefinelanguage{pcode}{
	language=none,
	numbers=left,
    stepnumber=1,
    basicstyle=\scriptsize\ttfamily,
    morekeywords={INT_SUB, INT_LESS, INT_SEXT, INT_MULT, INT_ADD, LOAD, PTRSUB, CBRANCH},
    keywordstyle=\color{blue},
    sensitive=false, 
    morecomment=[l]{//}, 
    morecomment=[s]{/*}{*/}, 
    morestring=[b]", 
} %
\lstdefinelanguage{none}{
  identifierstyle=
}

\lstdefinelanguage{ASM}{
    morekeywords={b, ble, blt, bne, bx, bl, ldr, str, push, pop, mov, add, sub},
    keywordstyle=\color{blue},
    sensitive=false, 
    morecomment=[l]{//}, 
    morecomment=[s]{/*}{*/}, 
    morestring=[b]", 
} %
\lstdefinelanguage{none}{
  identifierstyle=
}

\usepackage{subcaption}

\usepackage{xspace}

\usepackage{bytefield}
\usepackage{xcolor,colortbl}

\usepackage{hyperref}
\hypersetup{
    colorlinks = false,
    linkbordercolor = {white}
}

\newcommand{\frida}{\textsc{f\reflectbox{r}ida}\xspace}

\begin{document}
%
\title{ARIstoteles -- Dissecting Apple's Baseband Interface}
%
%
\author{Tobias Kr\"oll\inst{1} \and
Stephan Kleber\inst{2} \and
Frank Kargl\inst{2} \and
Matthias Hollick\inst{1} \and
Jiska Classen\inst{1}}
\authorrunning{Kr\"oll et al.}
%
\institute{TU Darmstadt, Secure Mobile Networking Lab, Germany
\email{tkroell,mhollick,jclassen@seemoo.de}\\ \and
Ulm University, Institute of Distributed Systems, Germany\\
\email{\{stephan.kleber,frank.kargl\}@uni-ulm.de}}

\maketitle              
{\let\thefootnote\relax\footnotetext{This is the accepted manuscript of a chapter published in Springer. The final authenticated version is available online at: \url{https://link.springer.com/chapter/10.1007/978-3-030-88418-5_7}}}
%
%

\begin{abstract}

Wireless chips and interfaces expose a substantial remote attack surface.
As of today, most cellular baseband security research is performed on the
\emph{Android} ecosystem, leaving a huge gap on \emph{Apple} devices.
With \emph{iOS} jailbreaks,
last-generation wireless chips become fairly accessible for performance and security research. Yet,
\emph{iPhones} were never intended to be used as a research platform,
and chips and interfaces are undocumented. One protocol to interface with such chips
is \ac{ARI}, which interacts with the central phone component \texttt{CommCenter} and multiple user-space daemons,
thereby posing a \ac{RCE} attack surface.
We are the first to reverse-engineer and fuzz-test the \ac{ARI} interface on \emph{iOS}.
Our \emph{Ghidra} scripts automatically generate a \emph{Wireshark} dissector, called 
 \emph{ARIstoteles}, by parsing closed-source \emph{iOS} libraries for this undocumented protocol.
Moreover, we compare the quality of the dissector to fully-automated approaches based on static trace analysis.
Finally, we fuzz the \ac{ARI} interface based on our reverse-engineering results.
The fuzzing results indicate that \ac{ARI} does not only lack public security research but also has not been well-tested by \emph{Apple}.
By releasing \emph{ARIstoteles} open-source, we also aim to facilitate similar research in the future.


\end{abstract}

%
%
%


\section{Introduction}

Any component on a mobile system that is reachable over the air poses a security risk.
On smartphones, the largest attack surface is functionality related to the cellular baseband---they
pose the longest wireless range and most complex codebase.
However, many wireless components are closed-source and undocumented, such as wireless firmware and
vendor-specific system extensions on all mobile platforms or wireless daemons on \emph{iOS}.
Thus, these complex components are often untested by public security research.

We focus on \emph{iOS} since it experienced significantly less public testing
than \emph{Android}.
Recent \emph{iPhones} come with basebands from either \emph{Qualcomm} or \emph{Intel}.
\emph{Intel} basebands are predominant on the European market and are at least shipped
in the \emph{iPhone 7}, \emph{8}, \emph{Xs}, \emph{11}, and \emph{SE2} models.
The same chip as in the \emph{iPhone 11} and \emph{SE2} is used by the \emph{iPad Air 2020},
but telephony functionality is restricted.
Moreover, the \emph{Apple Watch 6 GPS+Cellular} has a feature-reduced \emph{Intel}
baseband variant.
The management protocol for \emph{Qualcomm} chips is called \ac{QMI}, which open-source libraries support~\cite{libqmi}. On \emph{Android} devices, \emph{Intel} chips
use AT commands as control interface~\cite{guy2019}. Likely for better interoperability and a feature set closer
to \ac{QMI}, \emph{Apple}
introduced \acf{ARI}.
Security of the \ac{ARI} interface is unstudied, even though it resides in the heart of the \emph{iOS} baseband ecosystem.

Understanding \ac{ARI} enables baseband research on \emph{Apple} devices.
Via debug profiles~\cite{debugprofile}, it supports passive baseband analysis on non-jailbroken devices.
Moreover, on jailbroken devices, this knowledge can be used to control the baseband chip.
Basic protocol structure information is already sufficient to inject mutated \ac{ARI} payloads for fuzzing purposes.
However, more fine-grained control requires understanding all protocol internals in detail.
To establish the understanding, we reverse-engineered \emph{iOS} \ac{ARI} parsing libraries. We instrumented them with our
\emph{Ghidra} scripts to automatically extract type information to generate a \emph{Lua}-based
\emph{Wireshark} dissector~\cite{ghidra,wireshark},
which we name \emph{ARIstoteles}.
More precisely, the contributions of this paper are as follows:
\begin{itemize}
\item We are the first to publicly \textbf{reverse-engineer and document the \ac{ARI} protocol}, resulting in the \emph{Wireshark} dissector \emph{ARIstoteles}.
\item Our \emph{Ghidra} scripts generate \emph{ARIstoteles}, which \textbf{automatically detect new protocol fields} added in future \emph{iOS} updates.
\item We compare the \emph{Ghidra}-based dissector generation approach to fully automated dissector \textbf{generation based on static analysis of protocol traces}.
\item By fuzzing the \ac{ARI} interface in user space with \frida~\cite{frida}, we
	encounter various \textbf{parsing errors and crashes in \emph{iOS} binaries}.
\end{itemize}

Overall, we identify \num{42} unique crashes in \texttt{CommCenter} and \ac{ARI}-related parsing libraries.
Due to the central role of \texttt{CommCenter} within \emph{iOS}---the telephony daemon in the \emph{iPhone}---these
crashes caused by the fuzzer also affect \num{12} daemons communicating with \texttt{CommCenter} via \ac{XPC}.
We emphasize that we disclosed all fuzzing results to \emph{Apple}, who started patching in \emph{iOS 14.2} and added further patches in \emph{iOS 14.6}.
\emph{ARIstoteles} and crashing example payloads are available online.\footnote{\url{https://github.com/seemoo-lab/aristoteles}}

This paper is structured as follows.
\autoref{sec:background} explains the baseband security model and security research in this field.
Before diving into further details, \autoref{sec:ari} describes the basic \ac{ARI} packet structure, which helps understanding
why static traffic-based approaches are not optimal to dissect \ac{ARI} automatically in \autoref{sec:ml}. Thus, we create
\emph{Ghidra} scripts in \autoref{sec:reversing}, which extract protocol information from shared libraries. These scripts work across \num{10}
different \emph{iOS} versions. We use this protocol knowledge to fuzz \texttt{CommCenter} via \ac{ARI} in \autoref{sec:fuzzing} and 
show that the \emph{ARIstoteles} dissector can be used to understand these crashes. We conclude our results and outline the security impact in \autoref{sec:conclusion}.


\section{Background and Related Work}
\label{sec:background}

The following section provides background information on the \emph{iOS}
baseband architecture as well as previous security-related research in this area.

\tikzset{>=latex}
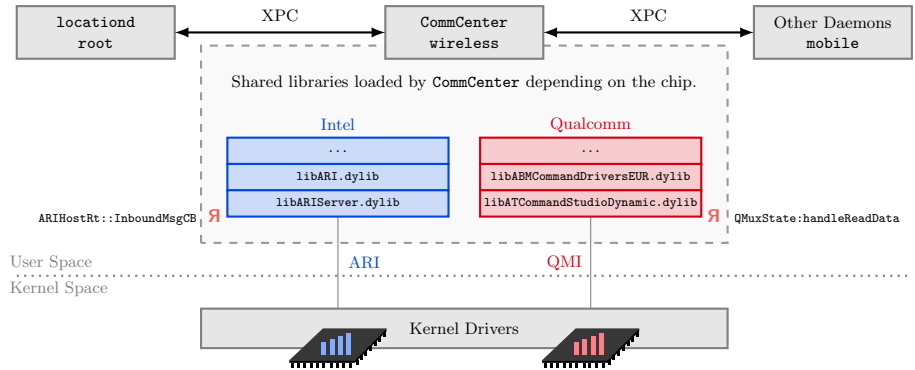
\begin{figure*}[b]
\vspace{-0.5em} 

	\center
	\begin{tikzpicture}[minimum height=0.55cm, scale=0.7, every node/.style={scale=0.7}, node distance=0.7cm] 
	
	
	\filldraw[fill=gray!5,draw=gray!80,thick,dashed,align=center](-3.5,-3.5) rectangle node (cc) {Shared libraries loaded by \texttt{CommCenter} depending on the chip. \\ \vspace{2cm}} ++(10,3.75);
	
	\filldraw[fill=gray!20,draw=gray,thick, align=center](0,0) rectangle node (cc) {\texttt{CommCenter}\\ \footnotesize{\texttt{wireless}}} ++(3,1);
	\filldraw[fill=gray!20,draw=gray,thick, align=center](-7,0) rectangle node (ld) {\texttt{locationd}\\ \footnotesize{\texttt{root}}} ++(3,1);
	\filldraw[fill=gray!20,draw=gray,thick, align=center](7,0) rectangle node (od) {Other Daemons\\ \footnotesize{\texttt{mobile}}} ++(3,1);
	
	\path[<->,thick] (-4,0.5) edge node[sloped,yshift=1em,align=center] {XPC} (0,0.5);
	\path[<->,thick] (3,0.5) edge node[sloped,yshift=1em,align=center] {XPC} (7,0.5);
	
	\draw[-,thin,gray] (-3+2.1,-3)--(-3+2.1,-5.5);
	\draw[-,thin,gray] (1.8+2.1,-3)--(1.8+2.1,-5.5);
	
	\node[align=center,color=darkblue] (l) at (-3+2.1,-1.25) {Intel};
	\filldraw[fill=darkblue!20,draw=darkblue,thick, align=center](-3,-2) rectangle node (o) {\scriptsize{\texttt{...}}} ++(4.2,0.5);
	\filldraw[fill=darkblue!20,draw=darkblue,thick, align=center](-3,-2.5) rectangle node (la) {\scriptsize{\texttt{libARI.dylib}}} ++(4.2,0.5);
	\filldraw[fill=darkblue!20,draw=darkblue,thick, align=center](-3,-3) rectangle node (las) {\scriptsize{\texttt{libARIServer.dylib}}} ++(4.2,0.5);	
	
	\node[align=center,color=darkred] (l) at (1.8+2.1,-1.25) {Qualcomm};
	\filldraw[fill=darkred!20,draw=darkred,thick, align=center](1.8,-2) rectangle node (o) {\scriptsize{\texttt{...}}} ++(4.2,0.5);
	\filldraw[fill=darkred!20,draw=darkred,thick, align=center](1.8,-2.5) rectangle node (la) {\scriptsize{\texttt{libABMCommandDriversEUR.dylib}}} ++(4.2,0.5);
	\filldraw[fill=darkred!20,draw=darkred,thick, align=center](1.8,-3) rectangle node (las) {\scriptsize{\texttt{libATCommandStudioDynamic.dylib}}} ++(4.2,0.5);
	
	\filldraw[fill=gray!20,draw=gray,thick,align=center](-3.5,-5.5) rectangle node (cc) {Kernel Drivers} ++(10,0.75);
	\node[inner sep=0pt] (chip) at (-3+2.1,-5.5) {\includegraphics[height=0.75cm]{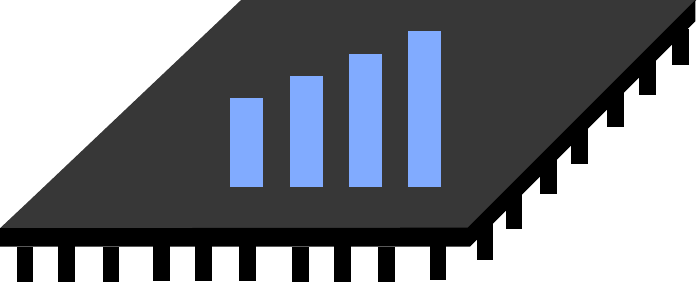}};
	\node[inner sep=0pt] (chip) at (1.8+2.1,-5.5) {\includegraphics[height=0.75cm]{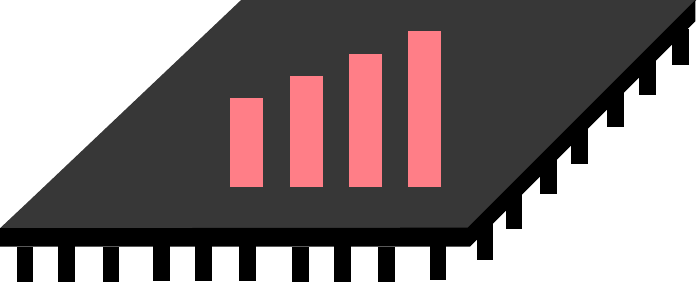}};
	

	\draw[-,thick,dotted,gray] (-7,-4.125)--(10,-4.125);

	\node[align=left,anchor=west,color=gray] (s) at (-7.25, -4.125+0.25) {User Space};
	\node[align=left,anchor=west,color=gray] (s) at (-7.25, -4.125-0.25) {Kernel Space};
	
	\node[align=left,color=darkblue] (s) at (-3+2.1+0.5, -4.125+0.25) {ARI};
	\node[align=left,color=darkred] (s) at (1.8+2.1-0.5, -4.125+0.25) {QMI};

	\node[align=left,anchor=east] (s) at (1.5-4.5, -3) {\scriptsize{\texttt{ARIHostRt::InboundMsgCB}} \hspace{0.1em} \normalsize\textcolor{frida}{\sffamily{\reflectbox{\textbf{R}}}}};
	\node[align=left,anchor=west] (s) at (1.5+4.5, -3) {\normalsize\textcolor{frida}{\sffamily{\reflectbox{\textbf{R}}}} \hspace{0.1em} \scriptsize{\texttt{QMuxState:handleReadData}}};

	\end{tikzpicture}

\vspace{-0.5em} 
\caption{Simplified baseband packet parsing architecture on \emph{iOS 12--14}.}
\label{fig:commcenter}
\end{figure*}

\subsection{Baseband Security Architecture}

When a wireless packet from a base station reaches a modern smartphone, it usually passes several data processing layers.
First, the baseband chip performs
initial processing. Then, it passes management information and data to the mobile operating system kernel running on the main processor. Usually, the kernel component
is as minimalistic as possible since it runs with high privileges. Hence, most protocol parsing is outsourced
to sandboxed daemons running in user space.

A concrete example of this separation on modern \emph{iOS} devices is
shown in \autoref{fig:commcenter}. The baseband chip processes data, which in turn passes information
to \texttt{CommCenter}. Due to intermediate data processing, some contents are not directly passed on to \emph{iOS}.
This abstraction layer might prevent some over-the-air attacks but can also introduce further mistakes.
For example, the baseband chip scans for nearby cells and measures their signal strength, and then further assembles this information into a list sent to \texttt{CommCenter}. Parsing issues could be on-chip, when directly processing cell information, but also within \texttt{CommCenter}, when parsing the list reported by the chip.
If abstraction by the chip prevents exploitation, an on-chip \ac{RCE} is required to escalate into \emph{iOS}.
Typically, wireless chips are optimized towards performance rather than security. Thus, baseband chips miss common
mitigation mechanisms and are a common exploitation target~\cite{nicointel,grant,basesafe,frankenstein}.
In the following, we assume that the attacker either already has code execution on the baseband chip (e.g., they can create an arbitrary list of cells with invalid entries and report it to \texttt{CommCenter}) or that data injected by an over-the-air attacker is not abstracted by the baseband chip (e.g., the cell identifiers are not further processed by the chip and directly forwarded to \texttt{CommCenter}).
Depending on the chip type, \emph{Intel} or \emph{Qualcomm}, \ac{ARI} or \ac{QMI} is the baseband management interface.

While this paper does not cover \emph{Android} in detail, the overall architecture is very similar.
However, \emph{Android} has a more diverse device landscape and enables vendors to add new wireless
chips. To this end, \emph{Android} defines a \ac{RIL}~\cite{ril}. Depending on the chip, a vendor-specific
library parses custom chip commands and translates them to \ac{RIL} calls.
Data between these components is passed as \emph{Binder} parcels. \emph{Binder} abstracts communication
between kernel and user space. Thus, it has been a popular research target for automated fuzzing
in the past~\cite{fans,badbinder}.

In contrast to \emph{Android}, baseband components on \emph{iOS} were not extensively researched.
In 2009, Mulliner and Miller approached \emph{iOS} fuzzing over the air by injecting SMS~\cite{blackhat2009iphone}.
Besides \ac{DoS}, they were able to gain code execution via an SMS parsing issue in \texttt{CommCenter}~\cite{26c3iphone}.
This bug was found on one of the very first \emph{iOS} releases---the first \emph{iPhone} was released in 2007---and
\emph{iOS 2} already had a \texttt{CommCenter} daemon but running as \texttt{root} without sandboxing.
Inspired by this bug, Silvanovich revisited SMS parsing on the significantly more hardened \emph{iOS 11} but could not find any issues~\cite{p0sms}.
SMS fuzzing is related to \ac{ARI} fuzzing, since \ac{ARI} also carries SMS payloads.
Apart from SMS parsing, we are not aware of any public security research revolving around \texttt{CommCenter}, which is surprising given that it poses a zero-click \ac{RCE} attack surface.

\subsection{Baseband Interface Analysis Options on iOS}

\emph{Apple} provides a baseband debug profile on the developer website~\cite{debugprofile}.
After installation on any \emph{iPhone}, it is possible to observe the baseband--\texttt{CommCenter} interface.
Not all capabilities of debug profiles are documented.
The documentation for the baseband debug profile only states the location of baseband chip crash logs,
which are stored in the undocumented \ac{ISTP} format.
Surprisingly, after installing
a baseband debug profile on an \emph{iPhone}---even if not jailbroken---all \ac{QMI} respectively \ac{ARI}
packets are printed to the system log as hexadecimal bytes.
Packets can be interpreted on the fly by parsing the log during runtime.

On jailbroken devices, it is furthermore possible to inject custom \ac{QMI} and \ac{ARI} payloads.
Silvanovich instrumented an undocumented \ac{XPC} interface for internal SMS testing~\cite{p0sms},
which injects SMS without passing the \ac{QMI} and \ac{ARI} parsing layer.
Instead of \ac{XPC} interfaces, we use \frida to directly hook into the \texttt{CommCenter} libraries responsible for parsing baseband messages~\cite{frida}.
Similar to a debugger, \frida can attach to a running process. It features support for modifying a process, injecting
payloads, and tracing program execution.
Thus, by hooking into a process with \frida, we are not restricted
to pre-defined \ac{XPC} interfaces and can even rewrite parts of the program code.

\subsection{iOS Shared Libraries}

Usually, \emph{iOS} shared libraries are
stored within one large \path{dyld_shared_cache} binary file. After attaching 
an \emph{iPhone} to \emph{Xcode} on \emph{mac\-OS} and setting it up for development, separated shared library files
are extracted to the folder \path{~/Library/Developer/Xcode/iOS DeviceSupport/}.
In contrast to most other \emph{iOS} binaries, shared libraries are meant to be
loaded and called by external processes. Thus, they export many function signatures,
making them easily accessible with tools like \emph{Ghidra} despite being closed-source.
The shared libraries handling \ac{ARI} are called \path{libARIServer.dylib} and \path{libARI.dylib}.

\path{libARIServer.dylib} is the last instance aware of protocol internals before passing
packets to kernel drivers. Thus, it is the perfect library for injecting arbitrary payloads. When sending information to the baseband, the function
\path{AriHostRt::SendRaw} is called, and the opposite direction uses the function \path{AriHostRt::InboundMsgCB}.
We will use these library functions for fuzzing in \autoref{sec:fuzzing}.
However, this library only performs a few final checks, such as checking sequence numbers, likely to prevent out-of-order
packets caused by multithreading.

\path{libARI.dylib} is closer to \path{CommCenter} and performs most of the
parsing and message abstraction. It contains human-readable names of all message groups
as well as definitions of most \ac{TLV} types. Hence, it is a great resource to automatically
generate a \emph{Wireshark} dissector, as shown in \autoref{sec:reversing}.


\section{Apple Remote Invocation Protocol}
\label{sec:ari}

\ac{ARI} is an \emph{Apple}-internal protocol without any public documentation.
While we are the first to analyze \ac{ARI}, it would be hard to follow our
reverse engineering process and problems encountered using different approaches
without a basic understanding of the protocol structure.
Technical details about how most of the information was obtained will follow in
\autoref{sec:reversing}.

\newcommand{\baselinealign}[1]{%
  \centering
  \strut\small#1%
}
\newcommand{\colorbitbox}[3]{%
  \sbox0{\bitbox{#2}{#3}}%
  \makebox[0pt][l]{\textcolor{#1}{\rule[-\dp0]{\wd0}{\ht0}}}%
  \bitbox{#2}{#3}%
}

\begin{figure}[b]
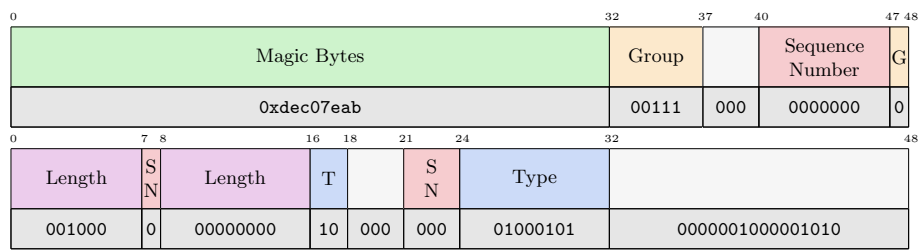

\adjustbox{width=1.0\textwidth}{
\begin{bytefield}[bitwidth=0.9333em, bitheight=3em]{48} 
\bitheader{0,32,37,40,47,48} \\
\bytefieldsetup{bitheight=3em}
\colorbitbox{darkgreen!20}{32}{Magic Bytes}
\colorbitbox{sorange!20}{5}{Group}
\colorbitbox{gray!7}{3}{}
\colorbitbox{darkred!20}{7}{Sequence Number}
\colorbitbox{sorange!20}{1}{G}
\\
\bytefieldsetup{bgcolor=gray!20,bitheight=2em}
\colorbitbox{darkgreen!20}{32}{\texttt{0xdec07eab}}
\colorbitbox{sorange!20}{5}{\texttt{00111}}
\colorbitbox{gray!20}{3}{\texttt{000}}
\colorbitbox{darkred!20}{7}{\texttt{0000000}}
\colorbitbox{sorange!20}{1}{\texttt{0}}
\\
\bitheader{0,7,8,16,18,21,24,32,48} \\
\bytefieldsetup{bitheight=3em}
\colorbitbox{darkpurple!20}{7}{Length}
\colorbitbox{darkred!20}{1}{S N}
\colorbitbox{darkpurple!20}{8}{Length}
\colorbitbox{blue!20}{2}{T}
\colorbitbox{gray!7}{3}{}
\colorbitbox{darkred!20}{3}{S\\ N}
\colorbitbox{blue!20}{8}{Type}
\colorbitbox{gray!7}{16}{}
\\
\bytefieldsetup{bgcolor=gray!20,bitheight=2em}
\colorbitbox{darkpurple!20}{7}{\texttt{001000}}
\colorbitbox{darkred!20}{1}{\texttt{0}}
\colorbitbox{darkpurple!20}{8}{\texttt{00000000}}
\colorbitbox{blue!20}{2}{\texttt{10}}
\colorbitbox{gray!7}{3}{\texttt{000}}
\colorbitbox{darkred!20}{3}{\texttt{000}}
\colorbitbox{blue!20}{8}{\texttt{01000101}}
\colorbitbox{gray!7}{16}{\texttt{0000001000001010}}
\\
\end{bytefield}
}
\vspace{-2em} 
\caption{ARI header format, \SI{12}{\byte} split into according bits with example values.}
\label{fig:ari-header}
\end{figure}

\begin{figure}[t]
\adjustbox{width=1.0\textwidth}{
\begin{bytefield}[bitwidth=0.9333em, bitheight=3em]{48} 
\bitheader{0,7,8,11,16,22,24,32,48} \\
\bytefieldsetup{bitheight=3em}
\colorbitbox{darkgreen!20}{7}{Type}
\colorbitbox{gray!7}{1}{}
\colorbitbox{sorange!20}{3}{Vers}
\colorbitbox{darkgreen!20}{5}{Type}
\colorbitbox{darkpurple!20}{6}{Length}
\colorbitbox{gray!7}{2}{}
\colorbitbox{darkpurple!20}{8}{Length}
\colorbitbox{darkred!20}{16}{Value}
\\
\bytefieldsetup{bgcolor=gray!20,bitheight=2em}
\colorbitbox{darkgreen!20}{7}{\texttt{0000010}}
\colorbitbox{gray!7}{1}{\texttt{0}}
\colorbitbox{sorange!20}{3}{\texttt{001}}
\colorbitbox{darkgreen!20}{5}{\texttt{00000}}
\colorbitbox{darkpurple!20}{6}{\texttt{000010}}
\colorbitbox{gray!7}{2}{\texttt{00}}
\colorbitbox{darkpurple!20}{8}{\texttt{00000000}}
\colorbitbox{darkred!20}{16}{\texttt{0000000000000000}}
\\
\end{bytefield}
}
\vspace{-2em} 
\caption{ARI TLV format with example payloads.}
\label{fig:ari-tlv}
\vspace{-1em}
\end{figure}

As a management protocol, \ac{ARI} only carries commands and information of a
more general nature without large amounts of data. For example, it signals incoming phone
calls---but the audio is sent over a different interface. Moreover, network traffic
is not handled by \ac{ARI}. Nonetheless, \ac{ARI} carries lightweight information besides
connection management, including SMS, location, and time.
Each \ac{ARI} packet has a fixed-length header with a group and type identifier.
The group defines the overall purpose, e.g., SMS processing, making calls, SIM access,
diagnostics, and more. Group types define further actions, such
as acknowledging an SMS, a request to send an SMS, or the delivery status of an SMS.
Furthermore, the header defines the remaining packet length, which is
filled with multiple \acf{TLV} structures. \acp{TLV} can be mandatory or optional.
The minimum packet length is \SI{12}{\byte} for packets that only have a header
and no \ac{TLV} payload. 

Since \ac{ARI} is for \emph{Apple}-internal use only, the protocol structure has grown historically. This shows in the
header format in \autoref{fig:ari-header}.
For example, the sequence
number was probably added to unused bits within three non-consecutive bytes.
This makes the header extremely hard to understand.

\acp{TLV} following the header 
are shown in \autoref{fig:ari-tlv}. In addition to a type, length, and value, they also
have a version number.
This packet structure still does not reflect protocol semantics.
However, it outlines that \ac{ARI} is rather complex and historically grown, making further security research challenging.


\section{Fully-Automated Protocol Dissection}
\label{sec:ml}

Manually analyzing large packet
traces and spotting patterns is a tedious task.
Despite being experienced with protocol reverse engineering, we failed to identify
most fields in the header manually in the beginning when we did not have the
information from \autoref{sec:ari}.
Thus, as a first step,
we use a heuristics-based segmentation only relying on traffic traces, which has been very successful on various other
protocols, including the proprietary \ac{AWDL} protocol~\cite{milan}. 
In this process, which is based on \textsc{nemesys} and \textsc{nemetyl}~\cite{nemesys,nemetyl}, we identify segment boundaries that are the basis to classify packet types.

We run both algorithms on a trace with \num{1000} packets.
The trace is collected from an \emph{iPhone} when making a phone call and receiving SMS.
The overall trace contains \num{2497} packets, and we reduce it by 
subsampling the recorded trace to \num{1000} messages to make the classification computationally feasible. For retaining most of the variance across the reduced amount of messages, we select the most dissimilar packets. We measure the dissimilarity by the number of uncommon byte sequences per message.
Subsampling reduces the groups from \num{22} to \num{20}. The by far most frequent group
is \path{net_cell} (labels obtained during later analysis) with \num{312} occurrences. Almost \SI{95}{\percent} of all messages are of \num{9} different groups.

\begin{figure}[b]
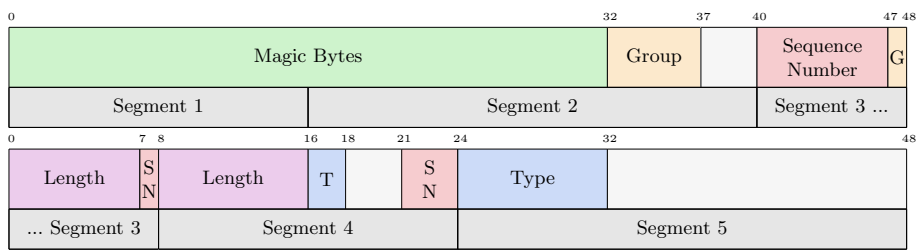

\vspace{-1em} 
\adjustbox{width=1.0\textwidth}{
\begin{bytefield}[bitwidth=0.9333em, bitheight=3em]{48} 
\bitheader{0,32,37,40,47,48} \\
\bytefieldsetup{bitheight=3em}
\colorbitbox{darkgreen!20}{32}{Magic Bytes}
\colorbitbox{sorange!20}{5}{Group}
\colorbitbox{gray!7}{3}{}
\colorbitbox{darkred!20}{7}{Sequence Number}
\colorbitbox{sorange!20}{1}{G}
\\
\bytefieldsetup{bgcolor=gray!20,bitheight=2em}
\bitbox{16}{Segment 1} &
\bitbox{24}{Segment 2} &
\bitbox{8}{Segment 3 ...} &
\\
\bitheader{0,7,8,16,18,21,24,32,48} \\
\bytefieldsetup{bitheight=3em}
\colorbitbox{darkpurple!20}{7}{Length}
\colorbitbox{darkred!20}{1}{S N}
\colorbitbox{darkpurple!20}{8}{Length}
\colorbitbox{blue!20}{2}{T}
\colorbitbox{gray!7}{3}{}
\colorbitbox{darkred!20}{3}{S\\ N}
\colorbitbox{blue!20}{8}{Type}
\colorbitbox{gray!7}{16}{}
\\
\bytefieldsetup{bgcolor=gray!20,bitheight=2em}

\bitbox{8}{... Segment 3} &
\bitbox{16}{Segment 4} &
\bitbox{24}{Segment 5} &
\\
\end{bytefield}
}
\vspace{-2em} 
\caption{Heuristic-based analysis results for the ARI header.}
\label{fig:ari-header-auto}
\end{figure}

\autoref{fig:ari-header-auto} shows the manually reverse-engineered header format on top
and the result of the \textsc{nemesys} segmentation on the bottom. With one exception in the magic
bytes, all segment boundaries are correct. However, the segmentation algorithm 
only tries to find boundaries on a byte level and not on a bit level. Moreover, the
segmentation algorithm is not aware of non-continuous segments.

It is important to note that
these two properties, bit-level and non-continuous segments, are scarce for most protocols,
with a few exceptions like flag fields that would not lead to such a high fragmentation.
While bit-level segmentation would be easy to integrate into the existing segmentation
algorithm, identifying split segments and grouping them into one seems hard to achieve.
Adding more freedom to segmentation also increases the likelihood of detecting additional false segment boundaries in general.
Especially the bit-fragmented fields spread over multiple offsets are particular to \ac{ARI} and very uncommon for other protocols. These disjoint short chunks of bits make it almost impossible to
apply statistical methods in static traffic analysis.

\begin{figure}[b]
\centering
\vspace{-1em} 
\includegraphics[width=0.8\textwidth]{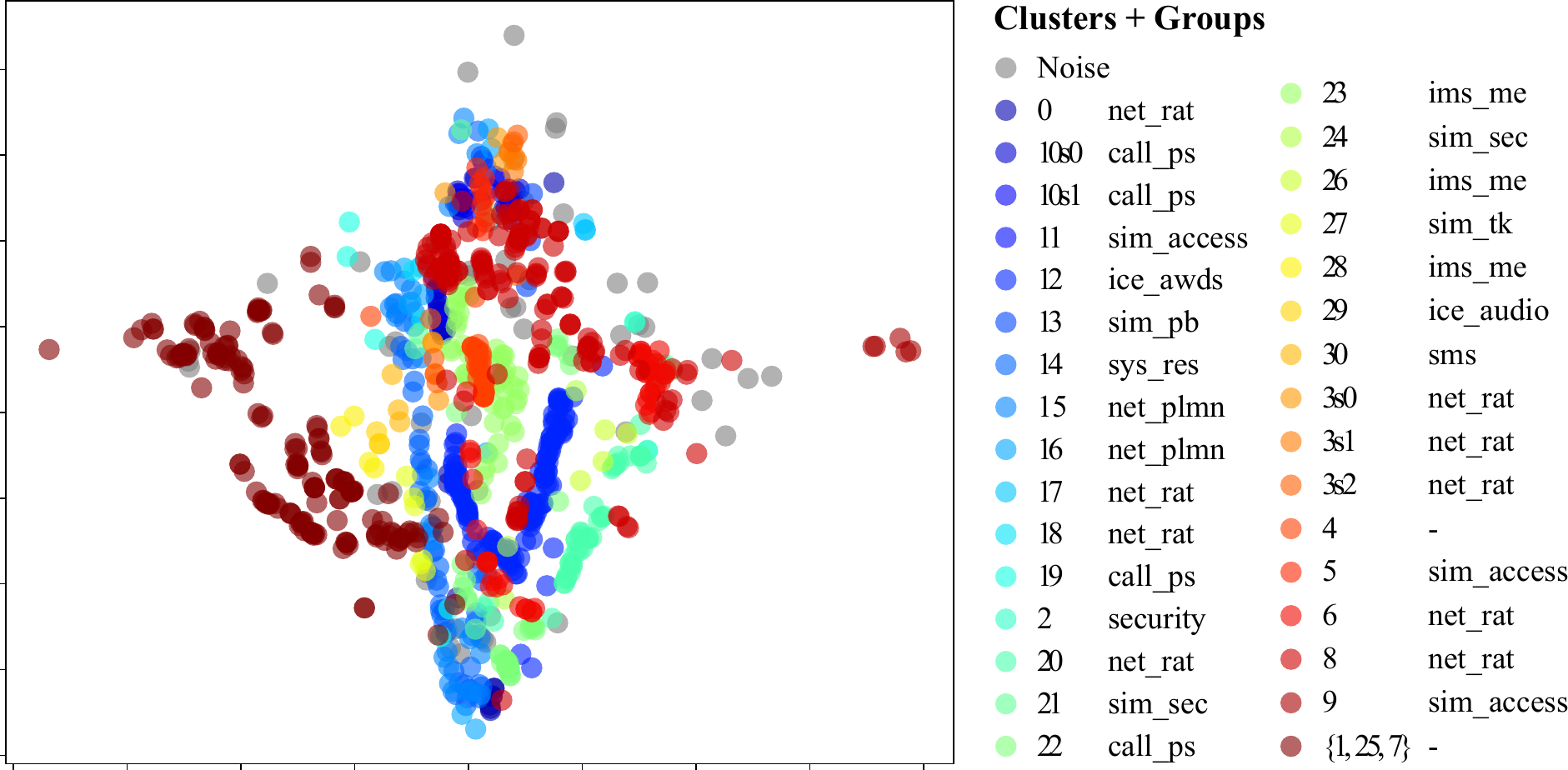}
\caption{Message type classification results showing multiple clusters.}
\label{fig:heuristic-msgtype}
\end{figure}

Next, we use \textsc{nemetyl} to identify the message type automatically. 
In contrast to a traditional packet parser, \textsc{nemetyl} does not rely
on detecting the group and type field in the header. Thus, even if the header
segmentation with \textsc{nemesys} fails, correctly segmenting parts of the
remaining packet and classifying them by the overall message similarity can be successful.
As shown in \autoref{fig:heuristic-msgtype}, \textsc{nemetyl} indeed
locates multiple clusters in the message trace. A unique color indicates a cluster.
To plot the dissimilarities of the
messages, their relative positions to each other are projected into a two-dimensional plane
using \ac{MDS}~\cite{mds}. The absolute positions in the plot are insignificant, therefore we omit any axis labels.
\textsc{nemetyl} identifies
\num{32} clusters. Only two of these clusters contain messages of multiple
groups, and the remaining \num{30} clusters belong to exactly one group, meaning
that the group classification is highly accurate.
Some of the correctly identified group clusters contain multiple message types.
Six group clusters contain two message types, and one group
contains four message types. Such mistakes can be explained with
similar packets belonging to different groups or types,
for example, an SMS in send versus receive direction.

We conclude that static traffic analysis shows a high potential to identify message types correctly.
Since \ac{ARI} is a protocol with uncommon properties, the analysis would
need further optimizations and very specific adaptions to work properly.
Some of the required changes are likely to only apply for \ac{ARI}.
Thus, instead of overfitting the heuristics, we continue with manual reverse-engineering and
automate these findings with \emph{Ghidra} scripts.
As future work, we plan to improve the static traffic analysis method with the insights gained in this use case to ease its application for other specialized unknown protocols.

\section{Automated Reverse-Engineering}
\label{sec:reversing}

Packet structure information is sufficient for basic security research, such
as generation-based fuzzing. However, to understand what exactly \ac{ARI} does and
utilize the baseband, additional information is required. \emph{iOS} needs some of
this information since it parses \ac{ARI} packets coming from the 
baseband and creates commands sent to the baseband.
Depending on the underlying chip, \emph{Qualcomm} or \emph{Intel},
\texttt{CommCenter} loads shared libraries for parsing \ac{QMI} or \ac{ARI}. In the following, we 
automatically analyze \ac{ARI} parsing in the \path{libARI.dylib} library to create a \emph{Lua}-based
dissector. First, we manually locate and reverse-engineer the central data structures and functions, which we then
parse automatically. The automated analysis even enables us to track protocol changes in \emph{iOS} updates.
While the initial parser generation needs manual work, it could be applied for other
proprietary protocols parsed by shared libraries. 
In addition to shared libraries, some \emph{macOS} binaries contain more symbols than \emph{iOS} binaries. Sometimes, \emph{Apple} accidentally releases a kernel without stripping symbols like in an early \emph{iOS 14 Beta} release.

\subsection{Group and TLV Definitions}

Message parsing starts at the function \path{AriMsg::AriMsg}. After checking that the
message starts with the magic bytes \texttt{0xdec07eab} and is at least \SI{12}{\byte} long,
it passes the message to \path{Ari::MsgDefById}.
This function assigns group information via a lookup in the 
array called \path{ARIMSGDEF_GROUPS}. This array contains \num{63} pointers to group
definitions on \emph{iOS 14.5}. Each group definition contains an array
of fixed-length message type objects. Likely, they were structures and arrays before compilation
and only contained data, no code.

\begin{figure}[b]
\adjustbox{width=1.0\textwidth}{
\begin{bytefield}[bitwidth=0.8em, bitheight=3em]{56}
\bitheader{0,4,8,16,24,32,48} \\
\bytefieldsetup{bgcolor=darkgreen!20}
\bitbox{4}{Group} &
\bitbox{4}{Type} &
\bitbox{8}{Padding} &
\bitbox{8}{Mandatory TLVs} &
\bitbox{8}{Available TLVs} &
\bitbox{16}{Unknown} &
\bitbox{8}{Name} \\
\bytefieldsetup{bgcolor=gray!20,bitheight=2em}
\bitbox{4}{\texttt{0x9}} &
\bitbox{4}{\texttt{0x101}} &
\bitbox{8}{\texttt{0x0}} &
\bitbox{8}{\texttt{0x1C6B4D468}} &
\bitbox{8}{\texttt{0x1D9507480}} &
\bitbox{16}{\texttt{0x0003...00}} &
\bitbox{8}{\texttt{0x1C6B847E0}} \\
\bytefieldsetup{bgcolor=white,bitheight=2em}
\bitbox{4}{\texttt{}} &
\bitbox{4}{\texttt{}} &
\bitbox{8}{\texttt{}} &
\bitbox{8}{\texttt{msg101\_mtlvs}} &
\bitbox{8}{\texttt{msg101\_tlvs}} &
\bitbox{16}{\texttt{}} &
\bitbox{8}{\texttt{IBINetSet...}} \\
\end{bytefield}
}
\vspace{-2em} 
\caption{First message object in \texttt{\_ARIMSGDEF\_GROUP09\_net\_cell}, defining the
message structure of \texttt{IBINetSetRadioSignalReportingConfiguration}.}
\label{fig:ghidra-group}
\end{figure}

\begin{figure}[t]
\adjustbox{width=1.0\textwidth}{
\begin{bytefield}[bitwidth=1.4em, bitheight=3em]{32}
\bitheader{0,4,8,16,24} \\
\bytefieldsetup{bgcolor=darkgreen!20}
\bitbox{4}{Index} &
\bitbox{4}{Type} &
\bitbox{8}{Padding} &
\bitbox{8}{Encoding} &
\bitbox{8}{Name} \\
\bytefieldsetup{bgcolor=gray!20,bitheight=2em}
\bitbox{4}{\texttt{0x1}} &
\bitbox{4}{\texttt{0x101}} &
\bitbox{8}{\texttt{0x0}} &
\bitbox{8}{\texttt{0x1DE283508}} &
\bitbox{8}{\texttt{0x1C6B666E5}} \\
\bytefieldsetup{bgcolor=white,bitheight=2em}
\bitbox{4}{\texttt{}} &
\bitbox{4}{\texttt{}} &
\bitbox{8}{\texttt{}} &
\bitbox{8}{\texttt{ARI\_IBIUInt32\_1\_CODEC}} &
\bitbox{8}{\texttt{nInstance\_t1}} \\
\end{bytefield}
}
\vspace{-2em} 
\caption{First TLV object in \texttt{msg101\_tlvs}, defining the
type structure of \texttt{nInstance\_t1}.}
\label{fig:ghidra-tlv}
\end{figure}

\autoref{fig:ghidra-group} shows one of these message type objects
in the \path{net_cell} group.
Each object contains the group number, which is the same as the array index.
This is followed by a unique message type identifier.
Moreover, each group definition contains three pointers: a list of mandatory
\acp{TLV}, a full list including optional \acp{TLV}, and a human-readable type name.

The first entry in the list of \acp{TLV} is shown in \autoref{fig:ghidra-tlv}.
Each \ac{TLV} definition starts with an incrementing index number and a
type identifier. Then, the actual definition follows, consisting of a pointer to the encoding
and the name.

We extract both structures, shown in \autoref{fig:ghidra-group} and \ref{fig:ghidra-tlv}, with
a \emph{Ghidra} script.
This way, we can access all information with a decompiler despite not having any source code.
The script starts at the symbol \path{ARIMSGDEF_GROUPS} in \path{libARI.dylib},
which is exported and, thus, easy to locate. Then, the script follows the pointers to the group and \ac{TLV}
definitions. Since the actual definition structure is removed during the compilation process, we apply
the reverse-engineered structural information from \autoref{fig:ghidra-group} and \ref{fig:ghidra-tlv}.
The script finishes information extraction when a \texttt{null} entry terminates a list.
Finally, all information is written into a \emph{Lua} table, which is included by the \emph{Wireshark}
dissector.

\subsection{Type Definitions}

As of \emph{iOS 14.5}, the shared \ac{ARI} library defines \num{290} \texttt{asString} methods.
The method definitions differ by the parameter type, which corresponds to an encoding.
An encoding specifies a human-readable format of a type. For example, a phone call can be disconnected due to \num{226} reasons,
represented by an integer, and the corresponding \texttt{asString} method transforms this number to a more descriptive reason.
The \texttt{asString} methods are code and, thus, do not follow the same simple memory structure
as the group and \ac{TLV} definitions. Moreover, likely due to compiler optimizations, each \texttt{asString} method looks
slightly different. During a manual comparison, we identify three variants:
\begin{enumerate}
\item A lookup table, where the input is used as index in a list of strings.
\item An if-not construct, where strings are set to a default value and changed if the default does not apply in a fall-through manner.
\item A switch-case construct containing early returns, if statements, and jumps.
\end{enumerate}

The first two variants account for \num{240} of \num{290} \texttt{asString} methods and can be automatically extracted with \emph{Ghidra}.
Due to the inconsistent structure of the third variant, we have to extract these definitions manually.
It would also be possible to emulate the \texttt{asString} methods. However, their input range is undefined, and without analyzing their code, emulation might miss some inputs.

\begin{figure}

\begin{lstlisting}[language=C, caption={\texttt{asString} method with a look-up table (variant 1).}, label=lst:asstring]
char* asString(IBIImsMEAudioEVSBandWidthType t) {  // at address 0x1e0d2ab38
  if (t - @\textcolor{darkred}{1}@ < @\textcolor{darkpurple}{8}@) {
    return *(char **)((long)&@\textcolor{sorange}{PTR\_s\_IBI\_IMS\_ME\_AUDIO\_BAND\_NB\_1e0d2ab38}@ + (long)(int)(t - 1) * 8);
  }
  return "???";
}
\end{lstlisting}
\vspace{-2em} 

\end{figure}

We show how to extract the first variant, but all scripts and the \emph{Wireshark} dissector are available online.
Each lookup table has a wrapper function like the one shown in \autoref{lst:asstring}. It optionally adds an offset to the type
prior to the lookup in the string list. If the value is undefined, the string \textcolor{darkgreen}{\texttt{???}} is returned.
Mapping strings requires locating the lookup table as well as the range of input values. To this end, we need to interpret 
\emph{Ghidra} \emph{P-Code} instructions, which are the internal architecture-independent code representation within \emph{Ghidra}.
On a more detailed level, the extraction script works by iterating over all
\emph{P-Code} operands of the \path{asString} function.

\begin{figure}
\centering
\includegraphics[width=0.8\textwidth]{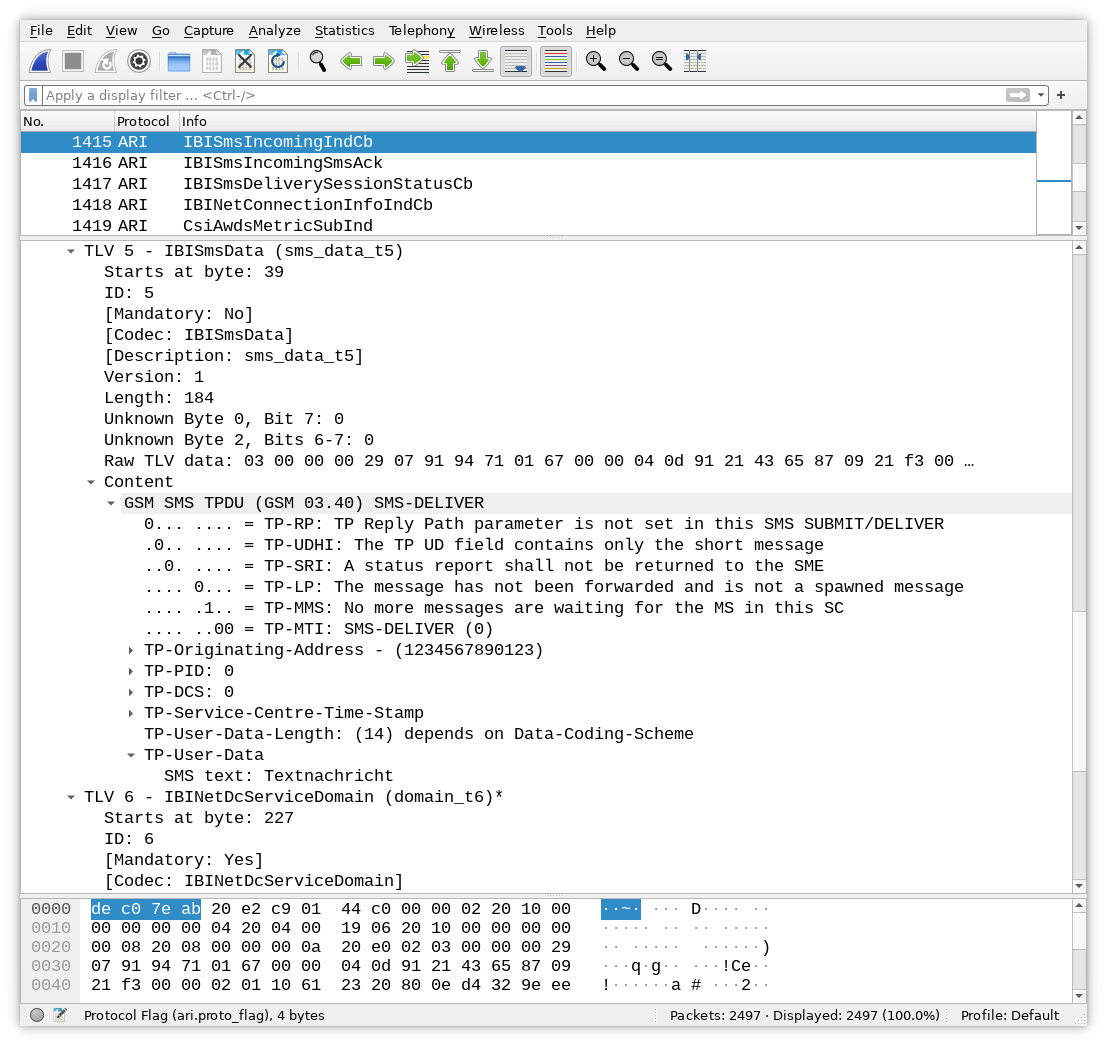}
\caption{\emph{ARIstoteles} dissecting an SMS.}
\label{fig:dissector-final}
\end{figure}

\subsection{Integrating Existing Dissectors}

\ac{ARI} embeds further formats that are well-known, such as SMS.
After adding these dissectors manually, the \emph{Wireshark} output
becomes even more readable. \autoref{fig:dissector-final} shows how
the resulting dissector looks after adding \ac{TLV} and type
definitions as well as SMS parsing.

\subsection{iOS Version Change Tracking}

Using the resulting \emph{Ghidra} scripts enables tracking changes in the \ac{ARI} protocol over
multiple \emph{iOS} versions. The results in \autoref{tab:iosversion} show that automated \emph{Wireshark}
dissector generation does not only save manual reverse-engineering overhead during initial dissector generation 
but with every \emph{iOS} update. This especially applies to \ac{TLV} types.
The similar numbers indicate that the scripts run flawless, independent of slight differences introduced by the
different compilation processes. Tracking changes within \ac{ARI} allows
determining if and when new features were added to the protocol and test these in specific.
Interestingly, the last \ac{ARI} update was applied in \emph{iOS 13.5}, released in parallel to 
the \emph{iPhone SE 2020}.

\begin{table}
\vspace{-1em} 
\caption{Extracted data structures for different \emph{iOS} versions.}\label{tab:iosversion}
\vspace{0.5em} 
\renewcommand{\arraystretch}{1.3} 
\centering
\scriptsize
\begin{tabular}{l|c|c|c}
\textbf{iOS Version} & \textbf{Types} & \textbf{\texttt{asString} (total)} & \textbf{\texttt{asString} (extracted)}\\
\hline
12.5.1 (16H22) & 783 & 290 & 249 \\
13.3 (17C54)   & 874 & 290 & 249\\
13.5 (17F75)   & 879 & 290 & 249\\
\hline
\textcolor{gray}{\emph{6 versions skipped}} &\textcolor{gray}{---} & \textcolor{gray}{---} & \textcolor{gray}{---}\\
\textcolor{gray}{14.5 (18E199a)} & \textcolor{gray}{879} & \textcolor{gray}{290} & \textcolor{gray}{249}\\

\end{tabular}
\end{table}


\section{Fuzzing}
\label{sec:fuzzing}

Based on the protocol knowledge, we can now inject messages into \texttt{CommCenter}.
We leverage this for smart fuzzing, compare \ac{ARI} results to \ac{QMI}, and outline how to
build high-performance \emph{iOS} fuzzers. Finally, we analyze these crashes.

\subsection{Initial Fuzzing Considerations}

There are only a few options for fuzzing closed-source \emph{iOS} daemons.
\emph{Corellium}, the most advanced commercial \emph{iOS} emulation, does not support cellular daemons and hardware as of May 2021~\cite{corellium}.
Open-source emulators can only boot parts of the kernel but not the entire system~\cite{ios-offensivecon}.
Thus, we need to fuzz \texttt{CommCenter} on a physical \emph{iPhone}. Non-emulated fuzzing has various
drawbacks~\cite{embeddedfuzzing}, such as non-scalability and no possibility to reset the system state. Nonetheless, running on physical
hardware can find more realistic bugs. Given these circumstances, we present the considerations that led to our setup for fuzzing \ac{ARI} messages in the following.

\begin{figure}[b]
\vspace{-1em} 
\begin{lstlisting}[caption={Injecting an ARI packet within the \texttt{CommCenter} process with \frida.}, label={lst:frida}]
// define hook for ARI messages chip (BB) -> iOS (AP)
var InboundMsgCb_addr = Module.getExportByName('libARIServer.dylib', '_ZN9AriHostRt12InboundMsgCBEPhm');
var InboundMsgCb = new NativeFunction(InboundMsgCb_addr, "int64", ["pointer", "int64"]);

Interceptor.attach(InboundMsgCb_addr, {  // optional hook customization
    onEnter: function(args) {
        // fix ARI sequence number, collect basic block coverage, etc.
}});

InboundMsgCb(payload, length);           // call the target once
\end{lstlisting}
\end{figure}

\paragraph{In-process Fuzzing on a Physical Device}
Our fuzzer is based on \frida, which injects a new thread into the target process. This means that a partial or complete fuzzer
is running within the \texttt{CommCenter} process.
\autoref{lst:frida} shows a minimal code example that injects a single message using the \emph{JavaScript} API.
The payload is injected into the target function \path{AriHostRt::InboundMsgCB} within the \path{libARIServer.dylib} shared library.
The example only calls the target function once. For fuzzing,
payloads would need to be generated and the function had to be called repeatedly. Moreover, we attach
a so-called interceptor to the target function, which can read and modify input parameters before execution.
This is necessary to keep track and adjust the sequence numbers since the physical baseband 
continues sending messages during the fuzzing process, resulting in out-of-order sequence numbers.
Additionally, the \frida stalker can trace executed functions or code blocks, enabling coverage collection~\cite{stalker}. Execution is traced
by dynamic binary rewriting during runtime and executing copies of the original code that have additional tracing instructions.

\paragraph{Payload Injection}
Payloads can either be generated (1) by an external device and sent to the target or (2) within the
\frida thread. The first option enables fast development and reusing existing fuzzing libraries without
adapting them to \frida.
If the target process \texttt{CommCenter} crashes, the external device is still aware of the last packet
sent prior to the crash. Furthermore, the external device might have larger storage for packet and crash logs.
However, the second option is significantly faster. We explore and compare both strategies in the next section.

\paragraph{Payload Mutation}
Creating new payloads requires a mutation strategy, which can be (1) coverage-based, (2) generation-based, or a combination of both.
Even without much knowledge about a protocol, the \frida
stalker can collect basic block coverage and use this as a metric to improve the quality of injected packets. Whenever a packet reaches a new basic block, it is added to the corpus.
However, collecting coverage
with \frida slows down the fuzzer. Another issue that arises is inconsistent coverage.
Most binary parsers that process files would always produce the same coverage for the same
input, and processing the current file does not depend on the previous file. Protocols
are very different and highly state-dependent.
For example, if a multi-part SMS
is injected, the last message part would increase the coverage, but injecting this message part again would not
increase the coverage. Thus, even injecting the same packet twice in a row can
lead to different coverages. Furthermore, processes have an inconsistent state
that is not solely controlled by packets injected by the fuzzer but configurations, external daemons, or user interaction like
enabling a mobile hotspot and dialing a number, again leading to inconsistent coverage.
The second option, which is generating new messages based on protocol knowledge, is significantly faster and does not
lead to any issues with inconsistent coverage. We also compare coverage-based to generation-based approaches.

\subsection{Building and Optimizing Fuzzers}

We compare five fuzzers with multiple injection and mutation strategies.
These variants lead to significantly different speeds, as we verified by
fuzzing a comparable SMS payload with the performance results presented in \autoref{tab:fuzzspeed}.
These fuzzers do not only differ in speed but also type of crashes they are able to find.

\begin{table}[b]
\caption{Fuzzing speed depending on the \frida setup on an \emph{iPhone 7} with an \emph{iOS 14.3 Beta} and \frida \emph{14.1.2}.}\label{tab:fuzzspeed}
\vspace{0.5em} 
\renewcommand{\arraystretch}{1.3} 
\centering
\scriptsize
\begin{tabular}{ll|r}
& \textbf{Speed Test Scenario} & \textbf{Fuzz Cases per Second} \\
\hline
\emph{(5a)} & Local \emph{JavaScript} SMS bit flipper & \num{17000} \\
\emph{(5b)} & Local \emph{JavaScript} SMS bit flipper with additional array copy & \num{11000}\\
\emph{(4)} & Local \emph{JavaScript} SMS bit flipper collecting basic block coverage & \num{250}\\
\emph{(2)} & External \emph{Python} SMS injection & \num{400}\\
\emph{(3b)} & External \emph{Python} SMS injection collecting basic block coverage & \num{100}\\
\emph{(3a)} & External \emph{ToothPicker} variant with \texttt{radamsa} mutator and coverage &\num{20}\\
\end{tabular}
\end{table}

The prior reverse-engineering of the protocol internals enables us to
 fix certain aspects of the message, such as the sequence number.
Moreover, we leave the magic bytes in the header intact, and have options
to only mutate the \acp{TLV}.

\paragraph{(1) Replacing Existing Messages}
This approach randomly changes a few bytes after a random amount of messages. Ideally, the target \emph{iPhone} has a SIM card installed and is actively used to
make phone calls, send messages, browse the Internet, and so on. The total amount of messages highly depends on user interaction, thus, we
do not list it in \autoref{tab:fuzzspeed}, but it is significantly slower than all other variants.
When setting a good ratio of mutated versus original messages, the \emph{iPhone} works almost normally,
and the fuzzer can reach states rather deep within the protocol. Using this method, we found various crashes that cannot be reproduced by replaying a message trace.
For example, one null pointer issue in the audio controller can only be triggered when the user previously initiated a phone call---which requires user interaction
and cannot be reproduced by injecting \ac{ARI} messages alone.

\paragraph{(2) External Message Generation}
In this approach, we collect a large message corpus from typical baseband interaction. Then, we replay this corpus, either ordered or unordered, and mutate
a subset of the messages based on the protocol structure.
 All messages are logged on the external device since some
crashes only occur on a packet sequence. This fuzzer reaches around \num{400} fuzz cases per second (fcps).

\paragraph{(3) External Coverage-based \texttt{radamsa} Mutation}
One of our previous projects, \emph{ToothPicker}, fuzzes the Bluetooth daemon on \emph{iOS}~\cite{toothpicker}.
We adapt it to fuzz \texttt{CommCenter} and observe a slightly
slower speed than for the Bluetooth daemon, around \SI{20}{fcps} \emph{(3a)}.
Note that when injecting payloads without the smart \texttt{radamsa} mutation strategy, this could be improved up to \SI{100}{fcps} \emph{(3b)}.
Even when initializing the \texttt{radamsa} mutator with different payloads, it tends to find the same crashes over and over again.
Usually, mutators are trying to fuzz crashes again to determine the underlying issue further. However, in the case of an in-process fuzzer based on \frida,
crashing \emph{CommCenter} also leads to a crash of the fuzzer. A \texttt{CommCenter} restart takes around \SIrange{20}{30}{\second}.
After a crash, we remove the payload that lead to the crash from the corpus and reinitialize the fuzzer by collecting coverage.
On a large corpus, this can take several minutes, further reducing the fuzzer's speed.

\paragraph{(4) Local Coverage-based \texttt{AFL++} Mutation}
Another option is to mutate the payloads with \texttt{AFL++} on the target device.
This is meanwhile supported by \emph{fpicker}~\cite{fpicker}, and we were allowed to use an early pre-release version.
Running locally leads to a significant speedup of around \SI{250}{fcps}, now primarily limited by the \frida stalker collecting
coverage. This fuzzer is \SI{10}{\times} faster than \emph{ToothPicker} 
while supporting a similar feature set.

\paragraph{(5) Local High-speed Mutation}
We test the upper limit of achievable fuzzing speed by randomly flipping bits within a provided message. This is not a powerful mutator but shows that the
fuzzing speed could be increased up to \SI{17000}{fcps} \emph{(5a)}.
Already very small changes, such as copying the payload array slightly more inefficient \emph{(5b)}, lead to a speed reduction, as shown in \autoref{tab:fuzzspeed}.

Overall, we found the rather naive second approach the best to uncover new crashes.
Mutating payloads based on a large corpus and keeping structural protocol information intact reaches both a comparably fast
fuzzing speed and high bug density. Approaches collecting coverage were rather slow and suffered from inconsistent coverage
information, thereby not compensating massive slowdown introduced by coverage collection.

\tikzset{>=latex}
\begin{figure*}[h]

	\center
	\begin{tikzpicture}[minimum height=0.55cm, scale=0.7, every node/.style={scale=0.7}, node distance=0.7cm] 
	

	\filldraw[fill=gray!20,draw=gray,thick, align=center](0,0) rectangle node (cc) {\texttt{CommCenter}} ++(3,1);	
	\node[align=left,anchor=east, color=darkblue] (s) at (0, 0.5) {42/14/4};
	\node[align=left,anchor=west, color=darkred] (s) at (3, 0.5) {9/3/0};
	
	\filldraw[fill=darkblue!20,draw=darkblue,thick, align=center](0,-0.5) rectangle node (las) {\scriptsize{Intel}} ++(1.5,0.5);	
	\filldraw[fill=darkred!20,draw=darkred,thick, align=center](1.5,-0.5) rectangle node (las) {\scriptsize{Qualcomm}} ++(1.5,0.5);

	\filldraw[fill=gray!5,draw=gray!80,thick, align=center](-6,0) rectangle node (ld) {\texttt{locationd}} ++(3,1);	
	\node[align=left,anchor=east, color=darkblue] (s) at (-6, 0.5) {5/1/0};
	\node[align=left,anchor=west, color=darkred] (s) at (-3, 0.5) {3/2/0};
	
	\filldraw[fill=gray!5,draw=gray!80,thick, align=center](-3,2) rectangle node (ld) {\texttt{sharingd}} ++(3,1);
	\node[align=left,anchor=west, color=darkred] (s) at (0, 2.5) {1/0/0};
	
	\filldraw[fill=gray!5,draw=gray!80,thick, align=center](-6,-3) rectangle node (ld) {\texttt{abm-helper}} ++(3,1);
	\node[align=left,anchor=west, color=darkred] (s) at (-3, -2.5) {1/1/0};
	
	\filldraw[fill=gray!5,draw=gray!80,thick, align=center](2.5,2) rectangle node (ld) {\texttt{awdd}} ++(2,1);
	\node[align=left,anchor=east, color=darkblue] (s) at (2.5, 2.5) {1/0/0};
	
	\filldraw[fill=gray!5,draw=gray!80,thick, align=center](6,-1.5) rectangle node (ld) {\texttt{imagent}} ++(3,1);
	\node[align=left,anchor=east, color=darkblue] (s) at (6, -1) {1/0/0};
	
	\filldraw[fill=gray!5,draw=gray!80,thick, align=center](-6.5,1.5) rectangle node (ld) {\texttt{misd}} ++(2,1);
	\node[align=left,anchor=east, color=darkblue] (s) at (-6.5, 2) {4/0/1};
	\node[align=left,anchor=west, color=darkred] (s) at (-4.5, 2) {2/0/1};
	
	\filldraw[fill=gray!5,draw=gray!80,thick, align=center](-6.8,-1.5) rectangle node (ld) {\texttt{gpsd}} ++(2,1);
	\node[align=left,anchor=east, color=darkblue] (s) at (-6.8, -1) {1/0/0};
	
	\filldraw[fill=gray!5,draw=gray!80,thick, align=center](6,1.5) rectangle node (ld) {\texttt{suggestd}} ++(3,1);
	\node[align=left,anchor=east, color=darkblue] (s) at (6, 2) {1/0/0};
	
	\filldraw[fill=gray!5,draw=gray!80,thick, align=center](5.5,0) rectangle node (ld) {\texttt{mediaserverd}} ++(3,1);
	\node[align=left,anchor=east, color=darkblue] (s) at (5.5, 0.5) {2/0/0};
	
	\filldraw[fill=gray!5,draw=gray!80,thick, align=center](4.6,-3) rectangle node (ld) {\texttt{WirelessRadioMangerd}} ++(4,1);	
	\node[align=left,anchor=east, color=darkblue] (s) at (4.6, -2.5) {1/1/1};
	
	\filldraw[fill=gray!5,draw=gray!80,thick, align=center](-3,-1.5) rectangle node (ld) {\texttt{abmlite}} ++(2.5,1);
	\node[align=left,anchor=east, color=darkblue] (s) at (-3, -1) {1/0/0};
	
	\filldraw[fill=gray!5,draw=gray!80,thick, align=center](0,-2.5) rectangle node (ld) {\texttt{photoanalysisd}} ++(3,1);
	\node[align=left,anchor=east, color=darkblue] (s) at (0, -2) {2/0/0};

	\node[align=left,anchor=west] (s) at (-6.9,-3.75) {Individual crashes found: Total / Replayable / Fixed in iOS 14.3. Colors indicate \textcolor{darkblue}{ARI} vs. \textcolor{darkred}{QMI} fuzzing.};
	\end{tikzpicture}

\vspace{-0.5em} 
\caption{Fuzzing results, also affecting various daemons due to XPC.}
\label{fig:ari-crashes}
\vspace{-1em} 
\end{figure*}

\subsection{Crash Evaluation}

Every time a process crashes, \emph{iOS} creates a crash log containing the exception type, the name of the crashed thread within the process, as well as a backtrace of each thread.
Thus, we can identify unique crashes even without coverage-based fuzzing. \autoref{fig:ari-crashes} shows all crashes found by fuzzing \ac{QMI} and \ac{ARI}, including
processes that communicate with \texttt{CommCenter} using \ac{XPC}. Even though \ac{QMI} has fewer bugs, components residing on top of the \ac{QMI} protocol parser and
in external daemons crash as well, sometimes similar as during \ac{ARI} fuzzing.

Due to state issues, not all crashes are replayable by injecting the same messages as during fuzzing. For example, we find a crash in the Mobile Internet Sharing Daemon \texttt{misd},
caused by invalid memory access to an address that looks like a configuration string. This bug frequently occurs  on \emph{Qualcomm} and \emph{Intel} \emph{iPhones}, meaning that it is
chip-independent. However, it cannot be reproduced by replaying packets.
\emph{Apple} rated the \texttt{misd} parsing issue as security-critical but did not assign a \ac{CVE} number since they discovered it in parallel to our research. This bug was present in \emph{iOS 13.7} and fixed in \emph{iOS 14.2}.

\begin{figure}[t]
\centering
\includegraphics[width=0.8\columnwidth]{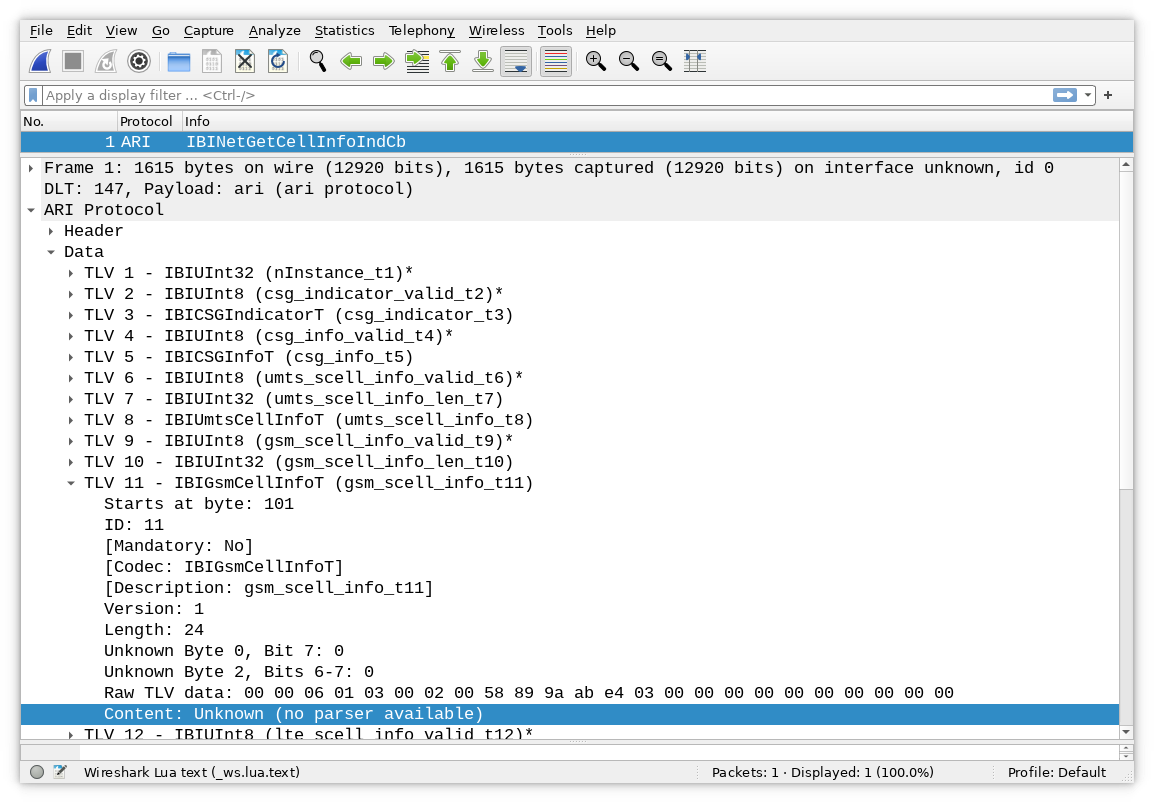}
\vspace{-1em} 
\caption{Crash within reported network cells during a scan.}
\label{fig:wireshark-crash}
\vspace{-1em} 
\end{figure}

Even when a bug is replayable, it is often hard to understand its root cause. In contrast to the shared library we were parsing in \autoref{sec:reversing}, \texttt{CommCenter} itself
does not contain exported symbols. Thus, we only know the crash location in a binary without function or variable names.
Reading code in the binary is further complicated by inaccuracies in the decompilation process, resulting in wrong pseudo-code output.
Understanding payloads significantly speeds up manual crash analysis.

We illustrate this with a crash that occurs in the \texttt{CellMonitor} thread. It accesses an invalid offset in the function \texttt{processGsmCellInfo}.
Converting the crashing packet into a \emph{Wireshark} dump and opening it
reveals which part of the payload causes the crash, as shown in \autoref{fig:wireshark-crash}, detailing the \texttt{IBIGsmCellInfoT} contents.
However, \emph{Apple} did not assign a \ac{CVE} number without a full proof of concept showing if this bug is actually exploitable beyond denial of service.

While not assigning high priority to most bugs found by our fuzzer, \emph{Apple} did indeed fix all replayable crashes as of \emph{iOS 14.6}. We assume that they were running our fuzzer, which we provided to them, and most likely with further internal analysis tools.

\section{Conclusion}
\label{sec:conclusion}

Proprietary and undocumented interfaces pose a huge security risk if not tested thoroughly by manufacturers.
Our fuzzing results let us assume that \emph{Apple} did not perform any fuzz testing
on \ac{ARI}. Given that this interface is reachable via wireless components, this is surprising.
We hope that \emph{Apple} will improve wireless interface security in future releases and
run---among other security testing---internal fuzzing campaigns.

The \emph{Wireshark} dissector we provide will be a helpful research tool also after \ac{ARI}
parsing issues got fixed within \emph{iOS}. It enables debugging baseband interactions on the fly
and provides insights on network operation, SIM interaction, and more. We hope that
the open-source release of \emph{ARIstoteles} will support wireless security research in general.
It is the first step towards a low-level cellular experimentation framework comparable to projects
that already exist for Bluetooth and Wi-Fi~\cite{mantz2019internalblue,nexmon}. Additionally, the
\emph{Ghidra} scripts will enable researchers to build similar dissectors based on shared library information or leaked symbols.

\newpage
\section*{Availability}
The \emph{ARIstoteles}
source code as well as the \emph{Ghidra} extraction scripts are available on \url{https://github.com/seemoo-lab/aristoteles}.
This repository also contains the \frida scripts that can record \ac{ARI} packets
during runtime and inject packets, including replayable crashing packets and packet sequences
found with the fuzzers.

%
%
%
\bibliographystyle{splncs04}
\bibliography{bibfile}

\end{document}